# Casimir radiation with Weyl semimetals

Yang Hu[1,2], Xiaohu Wu[1,3,#], Haotuo Liu[4], Mauro Antezza[5,6], and Xiuquan Huang[2,#]

[1]Shandong Institute of Advanced Technology, Jinan 250100, Shandong, P. R. China

[2]School of Power and Energy, Northwestern Polytechnical University, Xi'an 710072, Shaanxi, P. R. China

[3]Yangtze Laboratory, Wuhan 430205, Hubei, P. R. China

[4]Key Laboratory of Advanced Manufacturing and Intelligent Technology, Ministry of Education, Harbin University of Science and Technology, Harbin, 150080, P. R. China

[5]Laboratoire Charles Coulomb (L2C) UMR 5221 CNRS-Université de Montpellier, Montpellier F- 34095, France

[6]Institut Universitaire de France, 1 rue Descartes, Paris Cedex 05 F-75231, France

[#]Corresponding author: wuxiaohu@pku.org.cn (Xiaohu Wu);

xiuquan_huang@nwpu.edu.cn (Xiuquan Huang)

**Abstract**

When Casimir friction torque acts upon a rotated nanoparticle (NP), mechanical energy can be transformed into thermal energy, known as Casimir radiation, which significantly affects the thermal performance of nanoelectromechanical systems. In this work, we investigate Casimir radiation with nonreciprocal Weyl semimetals (WSM) NP levitated on a plate. WSM NP with inherent nonreciprocity has a radiative heat flux 27 times higher than NP with degenerate modes. The underlying physics is elucidated by the coupling and decoupling of the electromagnetic local density of states between nonreciprocal WSN NP and the plate in the near-field. The three-fold localized plasmon modes of WSM NP split into localized circular modes with strong gyrotropic response, which opens up new channels for Casimir radiation. This work provides a new method for nanoscale energy conversion in NP systems.

In 1948, Casimir proposed the "Casimir Effect" based on quantum fluctuation of electromagnetic waves. Nanoparticles (NPs) accelerated by Casimir friction torque emit heat to surroundings due to changes in the boundary conditions of the electromagnetic field, known as Casimir radiation [1, 2]. Pioneering work by A. Manjavacas et al. studied frictional torque acting on NPs rotating in a vacuum, demonstrating that Casimir friction transforms mechanical energy into thermal energy and produces NP heating [3, 4]. Recently, Deop-Ruano et al. investigated the near-field radiative heat transfer (NFRHT) and Casimir interactions between nanostructures, where the near-field radiative heat flux (NFRHF) can be increased, decreased, or even reversed due to rotation, making an important step toward the simultaneous exploration of energy conversion and angular momentum [5]. Previous studies have focused on the Casimir radiation of graphite NPs, which are abundant in interstellar dust. However, nonreciprocity —a fundamental property of our universe characterized by direction-dependent behavior—has not been considered in these investigations of Casimir radiation [6, 7].

In recent years, advances in material science and physics have sparked a revival of research on Casimir radiation. Weyl semimetals (WSM) are a distinctive class of anistropic materials that have garnered significant attention in thermal photonics owing to their unique electronic band structure and intriguing topological properties [8-10]. Hu et al. designed a high-performance thermal diode using the nonreciprocity of WSM, allowing NFRHT in a prior direction while being blocked in a reverse direction [11]. Guo et al. designed a thermal router that utilizes optical gyrotropy of WSM, which directs a NFRHF to a drain of choice by adjusting the momentum separation of Weyl nodes [12]. Ott et al. predict an anomalous thermal Hall effect mediated by photons in networks of WSM [13]. The inherent nonreciprocity and gyrotropy of WSM NP could promote the light-matter interaction and is expected to enhance and regulate Casimir radiation.

In this work, we investigate the Casimir radiation with rotated WSM NP in close vicinity to a plate with the same temperature. Inspired by previous works [14-17], a plate is placed in proximity to the NP to provide a huge local density of states (LDOS). We will analyze the case where the flat plates are hexagonal boron nitride (hBN) and WSM, respectively, and expect to observe an enhancement of Casimir radiation due to the coupling between NP and plate in the near-field.

The configuration of the Casimir radiation based on WSM NP is shown in **Fig. 1**(a). This work will provide two Casimir radiation enhancement schemes corresponding to the cases of WSM and hBN plates, respectively. The plate is semi-infinite and therefore we neglect the piezoelectric effects as a first approximation [18, 19]. The WSM NP is rotated around the $z$-axis with a frequency $\Omega = 10^9$ rad/s. The momentum separation **b** of WSM NP is along the $z$-axis. The temperature $T$ of the NP and the plate both are 300 K. In our calculations, the smallest gap separation considered is 50 nm. Additionally, the NP radius is 10 nm, which is significantly smaller than the gap distance. Such parameter settings are widely adopted in the study of NP-plate structures in NFRHT and the Casimir interactions. When the thermal wavelength and particle-plate distance is much larger than radius of nanoparticle, the NP can be regarded as pointlike source and the dipole approximation is valid [20, 21].

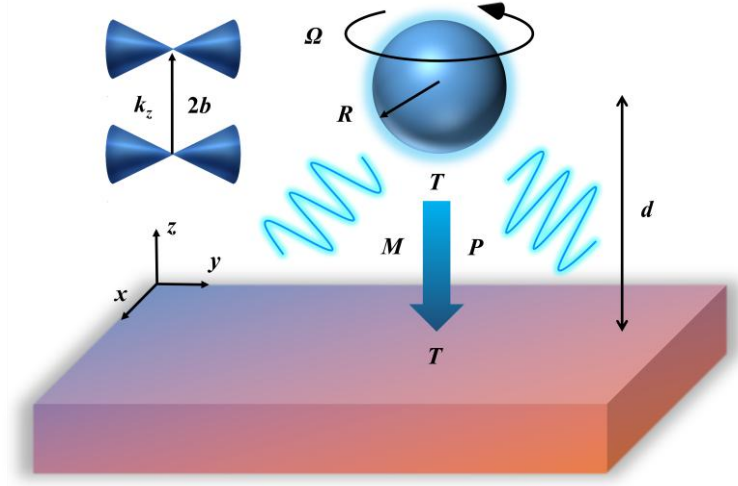

**Fig. 1** The configuration of the proposed system. The WSM NP with radius *R* is levitated above the plate along the *z*-axis with a gap distance of *d*. The WSM NP is rotated around the *z*-axis. The temperatures of NP and plate are both *T*. The NFRHT and Casimir torque are along the *z*-axis.

Casimir radiation is determined by the electric response of the material. Regarding the WSM, the time-reversal symmetry can be broken by splitting a Dirac point into a pair of Weyl nodes with opposite chirality separated by wavevector 2***b*** in momentum space. Furthermore, Weyl nodes alter the electromagnetic response, and the displacement electric field **D** can be expressed as [22, 23]:

$$\mathbf{D} = \varepsilon_d \mathbf{E} + \frac{ie^2}{4\pi^2 \hbar \omega}\left(-2b_0 \mathbf{B} + 2\mathbf{b} \times \mathbf{E}\right), \tag{1}$$

The specific meanings of parameters are in the supplementary materials. Dirac semimetals are typically considered isotropic when there are no external magnetic fields. Therefore, the diagonal elements of the permittivity tensor are assumed all $\varepsilon_d$. The term -$2b_0$**B** and 2**b**×**E** represent the chiral magnetic effect and optical gyrotropic response, respectively. We only consider $b_0 = 0$ where the Weyl nodes have the identical energy. The momentum-separation **b** of the Weyl nodes is an axial vector, similar to an internal magnetic field. We align the coordinates with **b** along the positive z-direction for NP (**b** = b**z**), resulting in the permittivity tensor of the WSM [34]:

$$\boldsymbol{\varepsilon} = \begin{bmatrix} \varepsilon_d & i\varepsilon_a & 0 \\ -i\varepsilon_a & \varepsilon_d & 0 \\ 0 & 0 & \varepsilon_d \end{bmatrix}, \quad (2)$$

The off-diagonal element:

$$\varepsilon_a = \frac{be^2}{2\pi^2 \varepsilon_0 \hbar \omega}. \quad (3)$$

When $b \neq 0$, $\varepsilon_a$ is nonzero, resulting in an asymmetric **ε**, potentially breaking Lorentz reciprocity. The value of $\varepsilon_a$ is comparable to $\varepsilon_d$ in the infrared region, indicating an exceptionally large and highly unusual gyrotropic response. The specific calculation parameters and the variation of dielectric constant with frequency are included in the supplementary materials. WSM is a very attractive and intricate material, and we focus here on the effect of its nonreciprocity on the Casimir radiation; for other material properties, i.e., titled Dirac cone, bandwidth, axion field, surface states and Weyl nodes etc., refer to the detailed discussion in [24-26].

The Casimir torque $M$ (integrate $M_\omega$ in frequency space) and NFRHT $P$ (integrate $P_\omega$ in frequency space) between NP and plate are calculated by the fluctuation-dissipation theorem [3, 4, 16, 34].

$$\begin{aligned} M_\omega(\omega) = & \frac{\hbar}{2\pi} \Big[ n_{np}(\omega-\Omega) - n_{pl}(\omega) \Big] \mathrm{Im}\Big[ \alpha_{xx}(\omega-\Omega) + \alpha_{yy}(\omega-\Omega) \Big] \mathrm{Im}\Big[ G_{xx}(\omega) + G_{yy}(\omega) \Big] \\ & + \frac{\hbar}{2\pi} \Big[ n_{np}(\omega-\Omega) - n_{pl}(\omega) \Big] \mathrm{Im}\Big[ \alpha_{xx}(\omega-\Omega) - \alpha_{yy}(\omega-\Omega) \Big] \mathrm{Re}\Big[ G_{xy}(\omega) - G_{yx}(\omega) \Big] \\ & + \frac{\hbar}{2\pi} \Big[ n_{np}(\omega-\Omega) - n_{pl}(\omega) \Big] \mathrm{Re}\Big[ \alpha_{xy}(\omega-\Omega) - \alpha_{yx}(\omega-\Omega) \Big] \mathrm{Im}\Big[ G_{xx}(\omega) - G_{yy}(\omega) \Big] \end{aligned} \quad (4)$$

$$P_\omega(\omega) = \frac{\omega\hbar}{2\pi}\left[n_{np}(\omega-\Omega)-n_{pl}(\omega)\right]\text{Im}\left[\alpha_{xx}(\omega-\Omega)+\alpha_{yy}(\omega-\Omega)\right]\text{Im}\left[G_{xx}(\omega)+G_{yy}(\omega)\right]$$
$$+\frac{\omega\hbar}{2\pi}\left[n_{np}(\omega-\Omega)-n_{pl}(\omega)\right]\text{Im}\left[\alpha_{xx}(\omega-\Omega)-\alpha_{yy}(\omega-\Omega)\right]\text{Re}\left[G_{xy}(\omega)-G_{yx}(\omega)\right]$$
$$+\frac{\omega\hbar}{2\pi}\left[n_{np}(\omega-\Omega)-n_{pl}(\omega)\right]\text{Re}\left[\alpha_{xy}(\omega-\Omega)-\alpha_{yx}(\omega-\Omega)\right]\text{Im}\left[G_{xx}(\omega)-G_{yy}(\omega)\right]$$
$$+\frac{\omega\hbar}{\pi}\left[n_{np}(\omega)-n_{pl}(\omega)\right]\text{Im}\left[\alpha_{zz}(\omega)\right]\text{Im}\left[G_{zz}(\omega)\right] \quad (5)$$

Where $n_{np}(\omega)$ and $n_{pl}(\omega)$ are the Bose-Einstein distribution of NP and plate. $\boldsymbol{\alpha} = 4\pi R^{2l+1}\frac{l(\boldsymbol{\varepsilon}-1\mathbf{I})}{l\boldsymbol{\varepsilon}+(l+1)\mathbf{I}}$ is the polarizability of WSM NP, $\mathbf{G}$ is the Green's function in the dipole-plate geometry. The detailed derivations and discussions are in the supplementary material.

The Casimir torque and NFRHT are significantly affected by the LDOS [14, 27, 28]. We try to couple the LDOS of the NP with the plate to enhance the Casimir torque and near-field radiative heat flux. The orientation-averaged LDOS of WSM NP is shown in **Fig. 2**(a), which can be obtained by the following equation [28, 29]:

$$\frac{\rho}{\rho_0} = 1 + \frac{1}{8\pi k_0^3}\sum_{l=1}^{\infty}(l+1)(2l+1)\frac{(\boldsymbol{\alpha}-\boldsymbol{\alpha}^*)/2i}{(R+\delta)^{2(l+2)}}, \quad (6)$$

The LDOS are proportional to the imaginary part of the polarizability. Through exploring the polarizability $\boldsymbol{\alpha}$ ($l=1$), the analytical equation for mainly resonance modes can be expressed as:

$$\det(\boldsymbol{\varepsilon}+2\mathbf{I}) = (\varepsilon_d+2)\left[(\varepsilon_d+2)^2-\varepsilon_a^2\right] = 0, \quad (7)$$

According to Eq. (7), it can be seen that when $b \neq 0$, such a 3-fold degenerate mode splits into three singly degenerate modes. The resonance modes appear when $(\varepsilon_d+2)=0$ or $(\varepsilon_d+2)^2-\varepsilon_a^2=0$ are satisfied. The former corresponds to the localized plasmon modes corresponding to the resonance of $\rho_{zz}$ at $\omega = 2.05\times10^{14}$ rad/s, while the latter corresponds to the localized circular modes corresponding to the resonances of $\rho_{xx}$ and $\rho_{xy}$, which distribute on both sides of the resonance of $\rho_{zz}$ at $\omega_{m = +1} = 1.65 \times 10^{14}$ rad/s and $\omega_{m = -1} = 2.55 \times 10^{14}$ rad/s. The localized circular modes are momentum separation dependent and related to the off-diagonal elements affected by gyrotropic optical responses. It is worth noting that the magnitude of the $\rho_{xy}$ has the opposite value of $\rho_{xx}$ at lower frequency when the momentum separation is along the positive z-axis.

The LDOS of the hBN and WSM plate is obtained by [30, 34]:

$$\frac{\rho}{\rho_0} = \int_{-\infty}^{\infty} \frac{dk_x}{2\pi} \int_{-\infty}^{\infty} \frac{dk_y}{2\pi} \frac{\omega}{\pi c^2} \operatorname{Im} \frac{i}{2\sqrt{k_0^2 - \kappa^2}} \times \left[ 4 + \frac{2\kappa^2}{k_0^2} \left( (r_{ss} + r_{pp}) e^{2i\sqrt{k_0^2 - \kappa^2} z} \right) \right], \tag{8}$$

As a hyperbolic material, hBN can excite hyperbolic phonon polaritons over a wide frequency range. It can be observed from **Fig. 2**(b) that the LDOS of the hBN plate can be significantly enhanced within the hyperbolic bands. Broadband enhancement makes it easy to couple with NP modes. For the WSM plate, when the direction of momentum separation is in the plane, the LDOS exhibits nonreciprocity with resonances at different frequencies.

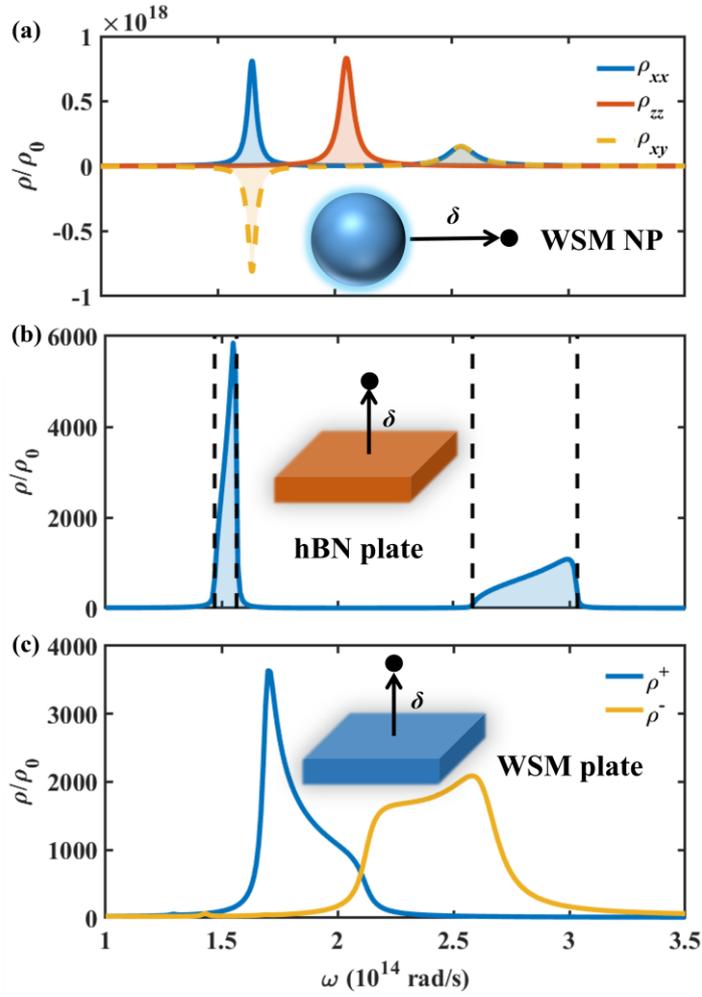

**Fig. 2** (a) The LDOS of WSM NP when $b = 1 \times 10^9$ m$^{-1}$. (b) The LDOS of a semi-infinite hBN plate. The black dashed lines present the transverse and longitudinal phonon frequencies. (c) The LDOS of a semi-infinite WSM plate for positive and negative directions of the wavevector when $b = 1 \times 10^9$ m$^{-1}$.

Due to the great tunability and nanoscale inhomogeneity of optical gyrotropy in WSM, the magnitude and direction of momentum separations **b** can be tuned easily by magnetization control [12, 31-33]. To clearly show the difference between WSM NP ($b \neq 0$) and isotropic WSM NP ($b = 0$), the enhancement ratio $\eta = M_{b \neq 0} / M_{b = 0}$ ($P_{b \neq 0} / P_{b = 0}$) is defined here. The enhancement ratio varying with momentum separations is shown in **Fig. 3**(a) when the plate is hBN. The shaded area represents the strong coupling region where the resonance of localized circular modes within the hyperbolic band is present. It can be seen that the enhancement ratio increases first and then decreases with the momentum separations. The enhancement ratio can be up to 66 for Casimir torque and 27 for NFRHF when $b = 1.3\times10^9$ m$^{-1}$; 31 for Casimir torque and 14 for NFRHF when $b = 1.5\times10^9$ m$^{-1}$, indicating that the coupling of the local circular modes in WSM NP and hyperbolic modes in hBN plate could enhance the Casimir radiation.

The localized circular modes are momentum separation dependent. We explore the polarizability varying with frequency with different momentum separations, as shown in **Figs**. 3(b)-(d). When $b = 0$, the off-diagonal elements of WSM permittivity are 0, and the imaginary part of $\alpha_{xx}$ and $\alpha_{zz}$ coincide. The WSM NP degenerates as isotropic Dirac semimetal, and the localized circular modes could not appear. When $b \neq 0$, the localized circular modes could be excited. The larger the value of momentum separation $b$, the further the localized circular mode resonance frequency is from the localized plasmon mode resonance frequency. The spin for quantum numbers $m = -1$ ($m = +1$) results in a blue (red) shift, making the splitting analogous to Zeeman splitting. When adjusting the value of momentum separations, the resonance of polarizability could be located within the hyperbolic band, promising enhancement of Casimir torque and radiative heat transfer.

The Casimir torque and NFRHT varying with angular frequency are shown in **Figs**. 3(e) and (f). For Casimir torque, when $b = 0$, the resonance at $\omega = 2.05\times10^{14}$ rad/s is attributed to the localized plasmon modes. The resonance within the shaded areas is attributed to the hyperbolic modes of the plate. When $b = 1.5\times10^9$ m$^{-1}$, the local circular modes coupling with the hyperbolic modes in type II hyperbolic band, leading to the significant enhancement of the Casimir torque. While $b = 3\times10^9$ m$^{-1}$, the resonant frequency of local circular modes are decoupling with hyperbolic bands, leading to the slight weakening of Casimir torque and NFRHF. The mechanism of NFRHT varying with frequency is similar to that of Casimir torque. We have performed additional calculations to benchmark the Casimir radiation for an isotropic NP made of a perfect metal (Ag) levitated above a hBN (depicted in purple). Similarly, we have also analyzed the Casimir radiation when a WSM NP (when $b = 1.5\times10^9$ m$^{-1}$) is levitated above an Ag plate (depicted in green). The results show that both the Casimir torque and the NFRHT are significantly weaker either the NP or the plate is made of Ag. This reduction is primarily due to the absence of resonant coupling between the NP and the plate, which plays a crucial role in enhancing Casimir radiation.

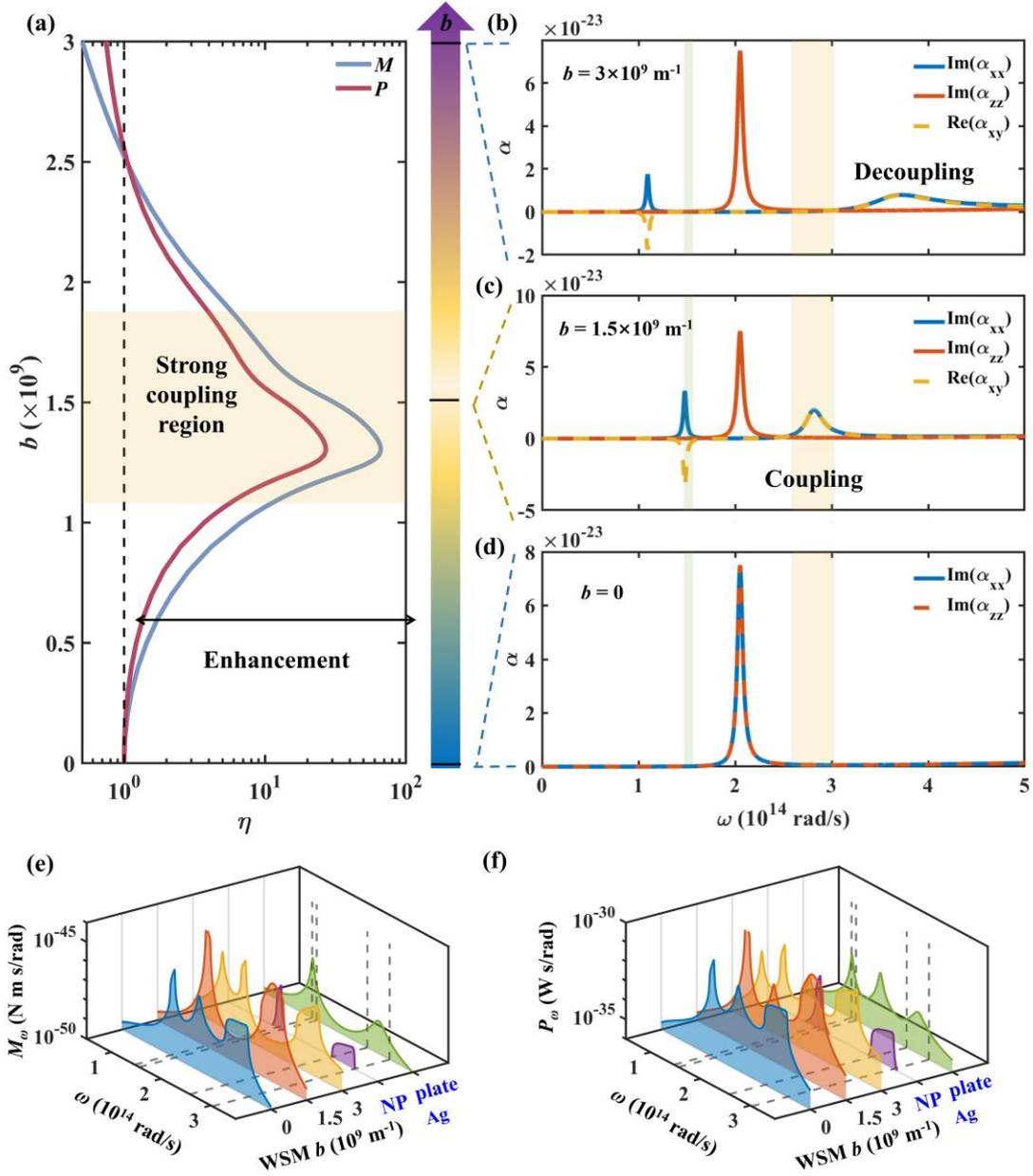

**Fig. 3** (a) The enhancement ratio varying with momentum separation. The main and off-diagonal elements of the polarizability tensor varying with frequency: (b) $b = 0$, (c) $b = 1.5\times10^9$ m$^{-1}$, and (d) $b = 3\times10^9$ m$^{-1}$. The shaded areas represent the hyperbolic bands of hBN. The spectral distribution of (e) Casimir torque (f) NFRHF. The black labels indicate the WSM NP and hBN plate structure. The blue labels indicate the Ag NP and hBN plate structure, and the WSM ($b = 1.5\times10^9$ m$^{-1}$) NP and Ag plates for comparison.

Next, we analyze the scenario where both NP and plate are WSM and the direction of momentum separation of WSM along the *y*-axis. The spectral Casimir torque and near-field radiative heat flux distribution are shown in **Figs. 4**(a) and (b). The spectral distribution for $+k_x$ and $-k_x$ is different, which can be understood by the reflection coefficient in **Figs**. 4(c) and (d). When momentum separation *b* is along the z-axis, the distribution of reflection coefficients at negative and positive wavevector space are symmetrical. The narrow and bright band at around $\omega = 2.05\times10^{14}$ rad/s is attributed to the excitation of surface plasmon polaritons (SPP). The resonance of SPP can well coincide with the resonance of localized plasmon modes of WSM NP. However, when momentum separation *b* is along the *y*-axis, the distribution of the reflection coefficients at negative and positive are asymmetric. The resonance of SPP split to a larger frequency at $-k_x$ space while to a smaller frequency at $+k_x$ space can be regarded as nonreciprocal SPP (NSPP). The resonance of NSPP can well coincide with the resonance of localized circular modes of WSM NP, as shown in **Fig**. 4(e).

The contribution of the electromagnetic model of the plate is embodied by the Green's function. Therefore, Green's function distributions are delved in wavevector space. When the momentum separation is along the *y*-direction, the reflection coefficients are unequal at the positive and negative *x*-direction. Therefore, Green's function distribution is asymmetric in the *x*-direction. When $\omega = 2.55 \times 10^{14}$ rad/s, Green's function is distributed at $-k_x$ wavevector, while at $+k_x$ wavevector when $\omega = 1.65 \times 10^{14}$ rad/s and $2.05 \times 10^{14}$ rad/s. The nonreciprocal distributions of Green's function lead to the nonreciprocal Casimir radiation property.

After analyzing the distribution of the reflection coefficients and Green's function, the mechanism of spectral heat flux resonance becomes clear. The resonances at $\omega = 1.65\times10^{14}$ rad/s and $\omega = 2.55\times10^{14}$ rad/s are attributed to the coupling of the localized circular modes of WSM NP and NSPP of the WSM plate. The resonances at $\omega = 1.65\times10^{14}$ rad/s are attributed to the coupling of the localized plasmon modes of WSM NP and bulk plasmon of the WSM plate. The Casimir radiation vary with gap distance are shown in Figs. 4(i) and (j) when the radius of WSM NP is 10 nm and 200 nm, respectively. For an NP with a 10 nm radius, Casimir radiation exhibits an expected exponential decay with increasing gap distance. However, for an NP with a 200 nm radius, the results reveal an oscillatory behavior. The observed sign oscillations in NFRHT and Casimir torque are related to NP resonances, suggesting that the direction and magnitude of heat flux and torque can be modulated by adjusting the NP-plate separation, and have a similar trend with the results in Ref. [16].

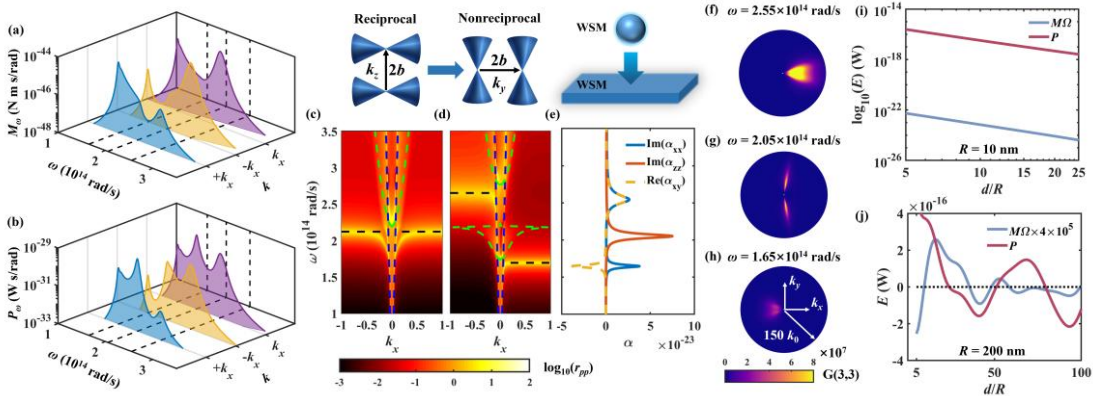

**Fig. 4** The spectral distribution of (a) Casimir torque and (b) NFRHF. The reflection coefficient varying with angular frequency and wavevector when the momentum separation is along (c) *z*-axis and (d) *y*-axis. The black lines are the linear dispersion relation in air (or vacuum). The magenta lines are the dispersion of surface plasmon polaritons. The green lines are the dispersion of bulk plasmon. The blue dotted lines represent the resonance frequencies of WSM NP. (e) The polarizability of WSM NP varying with frequency. The Green function varying with the dimensional wavevector when (f) $\omega$ = $2.55\times10^{14}$ rad/s, (g) $\omega$ = $2.05\times10^{14}$ rad/s, and (h) $\omega$ = $1.65\times10^{14}$ rad/s. $b = 1\times10^9$ m$^{-1}$. The Casimir torque and NFRHF varying with gap distance ($b = 1\times10^9$ m$^{-1}$ and along the $z$-axis), The radius of NP is (i) 10 nm and (j) 200 nm.

In summary, we investigate the Casimir radiation based on WSM NP. The NFRHF can be transferred from NP to plate without a temperature gradient with the rotation of nonreciprocal NP. The Casimir radiation can be enhanced by one order of magnitude compared with isotropic NP due to the coupling of localized circular modes of WSM NP and hyperbolic modes of the hBN plate. When the plate is WSM, the Casimir radiation exhibits nonreciprocal properties and can be improved by the coupling of localized circular modes and NSPP of the WSM plate. This work provides a new avenue to control the near-field energy based on nonreciprocity NP.

**Supporting Information**

Detail of dielectric function of WSM and hBN. Calculation of Casimir radiation using the fluctuation-dissipation theorem. Dipole approximation verification. The Weyl node number effect. Reflection coefficient calculation.

**Data Availability**

Data will be made available on reasonable request.

**Note**

The authors declare no competing financial interest

**Funding Sources**

This work is supported by the National Natural Science Foundation of China (52106099), the Natural Science Foundation of Shandong Province (ZR2022YQ57), the Taishan Scholars Program, Foundation of National Key Laboratory of Computational Physics, and China Scholarship Council program (202406290116).